\documentclass{article}

\usepackage{aimc2025}

\usepackage[utf8]{inputenc} 
\usepackage[T1]{fontenc}    
\usepackage{hyperref}       
\hypersetup{
  colorlinks   = true, 
  urlcolor     = blue, 
  linkcolor    = black, 
  citecolor   = black 
}
\usepackage{url}            
\usepackage{booktabs}       
\usepackage{amsfonts}       
\usepackage{nicefrac}       
\usepackage{microtype}      
\usepackage{graphicx}       
\usepackage[acronym]{glossaries}
\usepackage{subcaption}   
\usepackage{caption}      

\usepackage[utf8]{inputenc}
\usepackage[T1]{fontenc}
\usepackage{url}

\newacronym{eurorack}{Eurorack}{Eurorack}
\newacronym{cccb}{CCCB}{CCCB}
\newacronym{guell}{Palau Güell}{Palau Güell}

\newacronym{vce}{VCE}{Variational Cross-Examination}
\newacronym{gt}{GrooveTransformer}{GrooveTransformer}

\title{Exploring Situated Stabilities of a Rhythm Generation System through Variational Cross-Examination}

\author{%
  Błażej Kotowski\thanks{Equal contribution}\\
  Music Technology Group\\
  Universitat Pompeu Fabra\\
  Barcelona, Spain \\
  \texttt{blazej.kotowski@upf.edu} \\
  \And
  Nicholas Evans\footnotemark[1] \\
  Music Technology Group\\
  Universitat Pompeu Fabra\\
  Barcelona, Spain \\
  \texttt{nicholas.evans@upf.edu} \\
  \AND
  Behzad Haki  \\
  Music Technology Group\\
  Universitat Pompeu Fabra\\
  Barcelona, Spain \\
  \texttt{behzad.haki@upf.edu} \\
  \And
  Frederic Font \\
  Music Technology Group\\
  Universitat Pompeu Fabra\\
  Barcelona, Spain \\
  \texttt{frederic.font@upf.edu} \\
  \And
  Sergi Jordà\\
  Music Technology Group\\
  Universitat Pompeu Fabra\\
  Barcelona, Spain \\
  \texttt{sergi.jorda@upf.edu} \\
}

\begin{document}

\maketitle

\begin{abstract}
  This paper investigates GrooveTransformer, a real-time rhythm generation system, through the postphenomenological framework of Variational Cross-Examination (VCE). By reflecting on its deployment across three distinct artistic contexts, we identify three stabilities: an autonomous drum accompaniment generator, a rhythmic control voltage sequencer in Eurorack format, and a rhythm driver for a harmonic accompaniment system. The versatility of its applications was not an explicit goal from the outset of the project. Thus, we ask: how did this multistability emerge? Through VCE, we identify three key contributors to its emergence: the affordances of system invariants, the interdisciplinary collaboration, and the situated nature of its development. We conclude by reflecting on the viability of VCE as a descriptive and analytical method for Digital Musical Instrument (DMI) design, emphasizing its value in uncovering how technologies mediate, co-shape, and are co-shaped by users and contexts.

\end{abstract}





\section{Introduction}\label{sec:intro}

\begin{figure}[ht]
    \centering
    \begin{subfigure}[b]{0.44\textwidth}
        \centering
        \includegraphics[width=\textwidth]{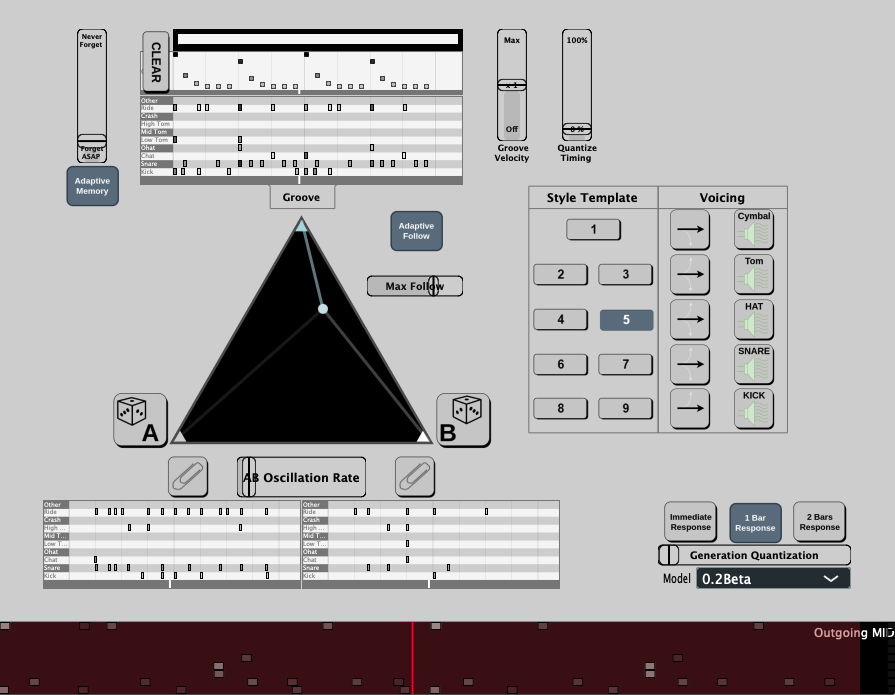} 
    \end{subfigure}
    \hspace{1em}
    \begin{subfigure}[b]{0.49\textwidth}
        \centering
        \includegraphics[width=\textwidth]{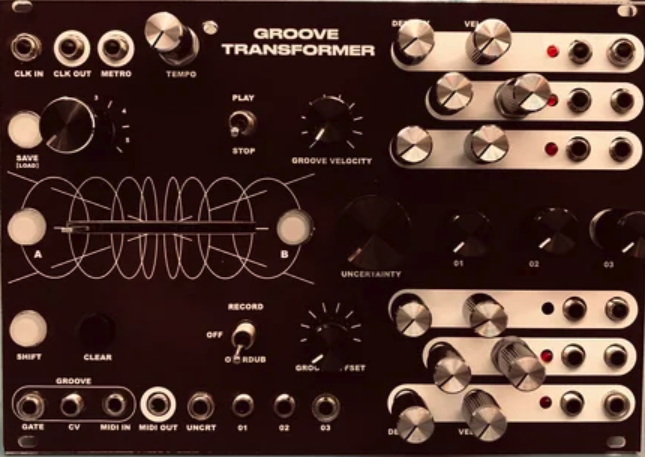} 
    \end{subfigure}
    \caption{VST plugin (left) and Eurorack (right) implementations of \acrshort{gt}.}
    \label{fig:soft_euro}
\end{figure}

The beginning of 2022 marked the start of two concurrent projects. The first, developed in collaboration with Raül Refree\footnote{https://en.wikipedia.org/wiki/Raül\_Refree}, aimed to design an autonomous rhythm generator for real-time  improvisation with minimal performer intervention; meanwhile, the second focused on developing a highly interactive, hardware-based generative rhythm sequencer. Although the two initiatives were driven by distinct objectives, their parallel development facilitated a productive exchange of ideas. This reciprocal influence ultimately culminated in the creation of a unified system, \acrshort{gt} \citep{GT_NIME, BEHZAD_THESIS}, which underlies both projects and integrates the core functions required by each (see Figure \ref{fig:soft_euro}).

\begin{figure}[ht]
  \centering
  \includegraphics[width=\linewidth]{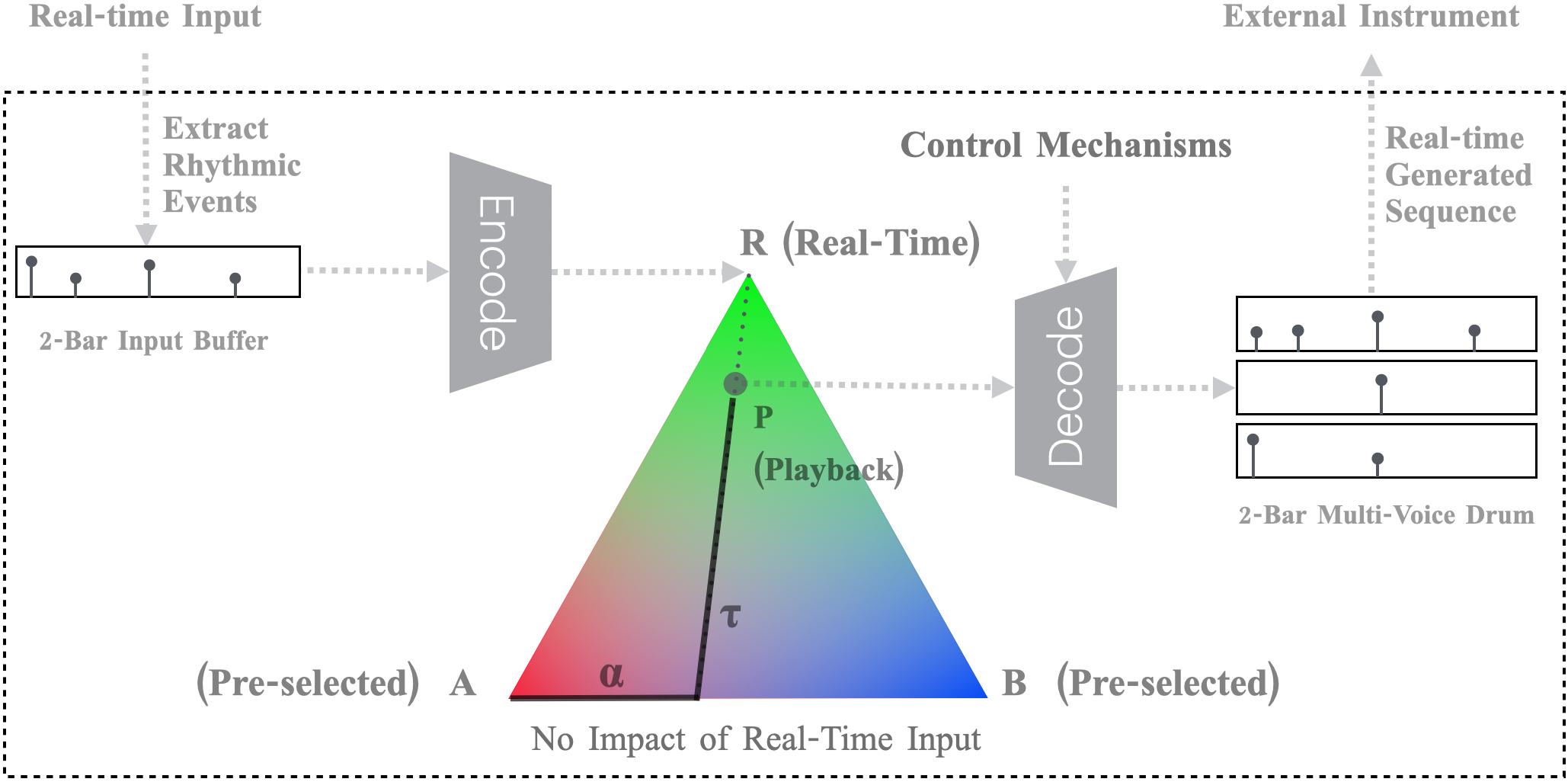}
  \caption{Conceptual design of \acrshort{gt}. The system operates within a defined subspace of the model’s latent space, bounded by three reference patterns: $A$, $B$, and $R$. Patterns $A$ and $B$ are pre-selected and static while Pattern $R$ is a dynamic pattern generated in real-time from rhythmic features extracted from the performer’s input. Positioning the playback point ($P$) results in bilinear interpolation along the $\alpha$ and $\tau$ dimensions and allows for the generation of intermediary patterns that blend characteristics of the three reference points. Navigating the space toward the top of the triangle increases the influence of the real-time performance ($R$), while navigating closer to the base reduces its impact in favor of the static patterns ($A$ and $B$). }\label{fig:triangular_area}
\end{figure}

At the core of \acrshort{gt} is a variational generative model trained to encode sequences of rhythmic onsets—including velocity and micro-timing—into a latent distribution from which multi-voice drum patterns are generated. This model underpins the system’s primary conceptual framework, as illustrated in Figure~\ref{fig:triangular_area}. 

\begin{figure}[ht]
    \centering
    \begin{subfigure}[b]{0.34\textwidth}
        \centering
        \includegraphics[width=\textwidth]{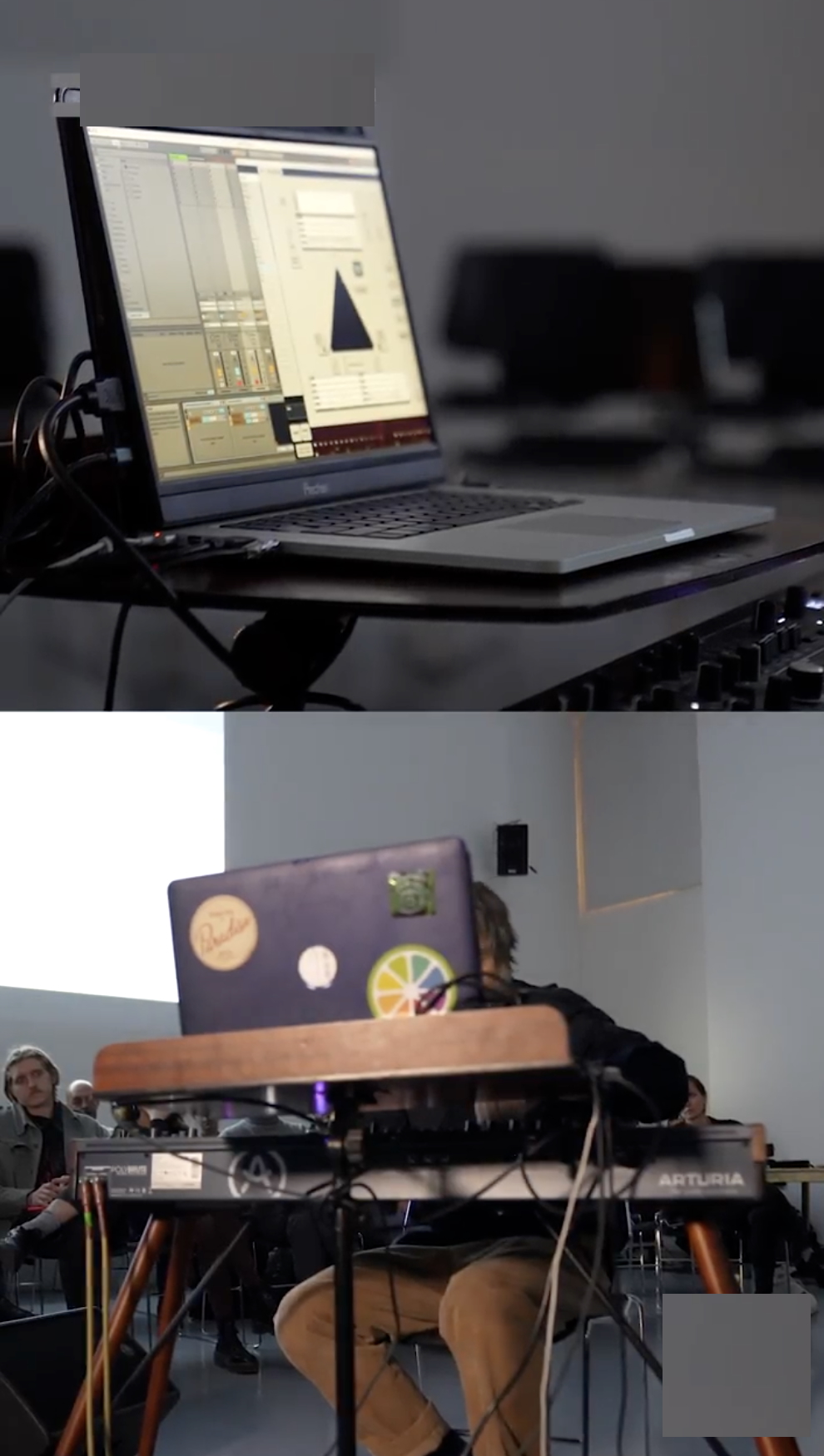} 
    \end{subfigure}
    \hspace{2em}
    \begin{subfigure}[b]{0.34\textwidth}
        \centering
        \includegraphics[width=\textwidth]{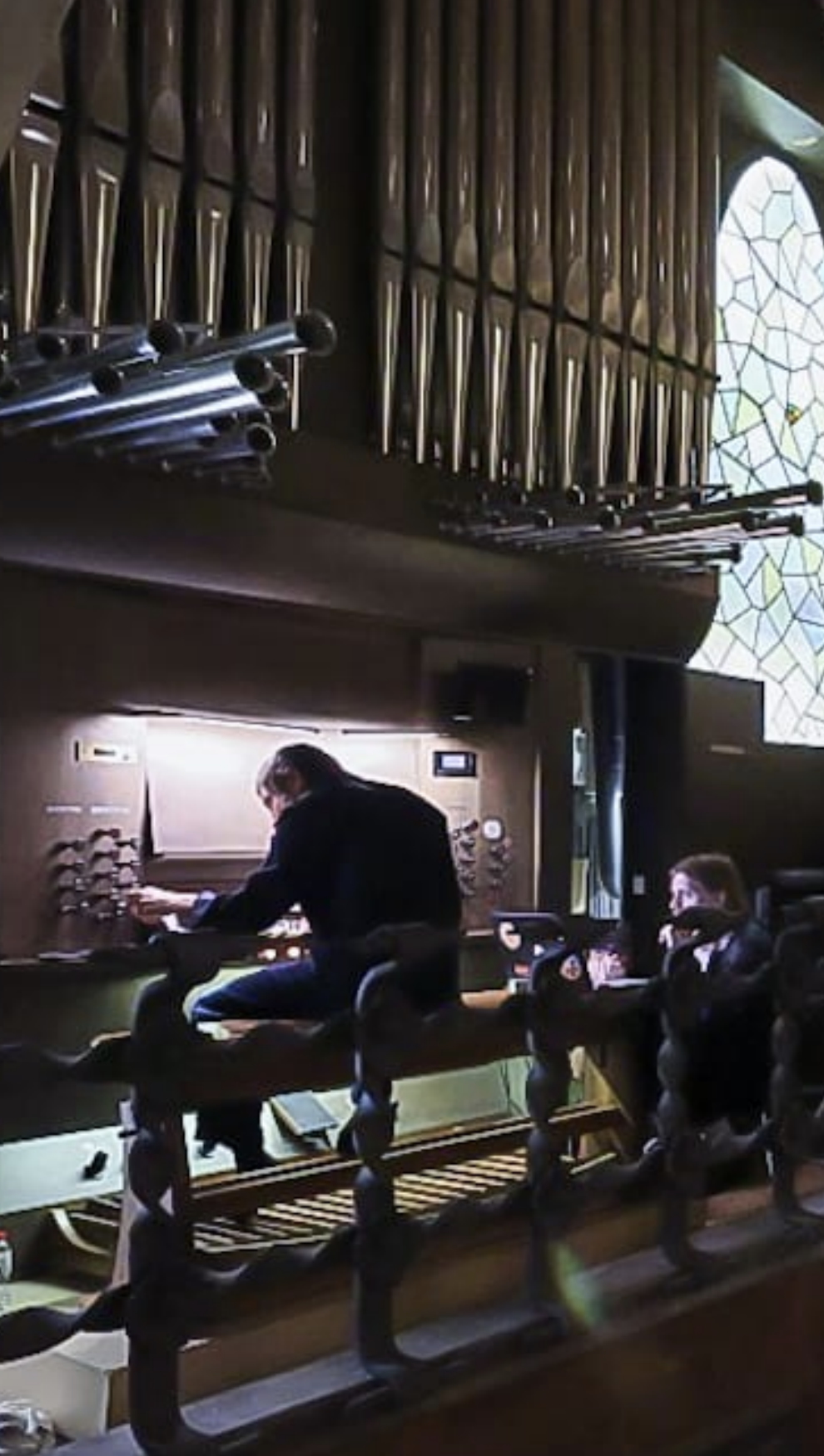} 
    \end{subfigure}
    \caption{Refree performances at \acrshort{cccb} (left) and \acrshort{guell} (right)}
    \label{fig:cccb_guell}
\end{figure}

\acrshort{gt} was showcased in two distinct contexts fulfilling two distinct roles. First, during a live performance at \acrshort{cccb} (Figure \ref{fig:cccb_guell}), the \acrshort{gt} VST functioned as an autonomous rhythm generator, providing real-time drum accompaniment to Refree’s improvisation on a keyboard synthesizer. Second, the hardware implementation—devloped as a Eurorack module—was presented at Sónar+D 2023, where its capabilities as a hands-on generative rhythm instrument were highlighted in a dedicated Eurorack showcase.

Following these events, a site-specific organ performance \citep{evans2025repurposing} was realized in collaboration with Refree at \acrshort{guell} (Figure \ref{fig:cccb_guell}). In this performance, Refree performed on the venue’s main MIDI-enabled organ console, while a generative system, designed specifically for this context, provided accompaniment on two portable, secondary MIDI-enabled organs. This generative accompaniment system, controlled by a second performer, integrated the \acrshort{gt} VST with a Markov-based pitch and note-duration generation engine, trained in real-time on Refree's performance. This configuration resulted in an autonomous layer that provided dynamic accompaniment on the portable organs, responding to the performance on the main organ.

The three distinct contexts in which the system was deployed demonstrate the flexibility of its design within noticeably different applications.\footnote{Video recordings available at \textbf{\url{https://situatedstabilities-aimc2025.github.io/}}} However, this versatility was not an explicit goal at the outset of the project. Instead, it gradually emerged through its open-ended development across multiple situated\footnote{As \citeauthor{verbeekWhatThingsPhilosophical2005} explains, situatedness talks about the intertwining of the humans, technical objects, and concrete contexts they find themselves in \citep[p.~33]{verbeekWhatThingsPhilosophical2005}.} settings. This raised a question for us: how did such different uses arise, especially without a specific methodology aimed at versatility? While the technical core of the system remained largely unchanged, its roles and meanings shifted significantly between contexts.

To examine this more closely, we invited a third researcher, who focuses on postphenomenological approaches to human-technology relations, to join the team. Together, we undertook a reflective inquiry into the ways in which the system was reconfigured through use. Rather than evaluating success or failure, we aimed to understand how its functions emerged through entanglements with specific practices, contexts, and interpretations.

The remainder of this paper is structured as follows. Section~\ref{sec:background} introduces the theoretical framework of postphenomenology and outlines the method of \acrfull{vce}, which guides our analysis. Section~\ref{sec:analysis} presents three distinct configurations of the system and examines them through three analytical lenses. Section~\ref{sec:reflections} reflects on these findings and discusses broader implications for DMI design. Finally, Section~\ref{sec:conclusions} offers concluding thoughts and directions for future work.



\section{Background}\label{sec:background}

Technologies resist fixed interpretations; rather, their roles, meanings, and effects emerge through use, situated within concrete practices and relations. Postphenomenology offers a framework for understanding this interpretive flexibility, not by analyzing technologies in isolation, but by studying how they mediate human-world relations \citep{verbeekTheoryTechnologicalMediation2015}.

In music performance, this mediation contributes largely to the notion of creative possibility. Rather than evaluating the tools in isolation, we can ask: how do they reorganize this creative possibility? What modes of interaction do they afford, constrain, or invite? Tools do not carry stable meanings, but shift according to how they are interpreted. The notion of \emph{multistability}, central to postphenomenology, explains this condition: technologies resist singular interpretations and stabilize differently depending on their context of use, user habits, and environmental affordances \citep{ihdeExperimentalPhenomenologyMultistabilities2012}. \citeauthor{ihdePostphenomenologyTechnosciencePeking2009} recalls his experience of being stabbed in the hand with a pen as a teen as a good example of how a single technology can be interpreted in multiple ways: a pen, most commonly used as a writing tool, in another context can double as a weapon. He devises a method of Variational Analysis (VA), which involves imaginative brainstorming of potential stabilities of specific technologies \citep{ihdePostphenomenologyTechnosciencePeking2009}.

While VA can be valuable in understanding the range of possible human-technology relations that a specific object might evoke, Rosenberger—a defining figure in contemporary postphenomenology—sees the potential of improving the depth of the analysis. He sees VA as a starting point for \acrfull{vce}, a method in which the identified stabilities are contrasted against one another \citep{rosenbergerVariationalCrossexaminationMethod2020}. This additional step in analysis helps to identify the meaningful differences between the stabilities, uncovering the underlying structure that determines which stabilities are possible. An important aspect of multistability is that while technologies allow for multiple interpretations, these are not unlimited, some interpretations are simply not possible. The emergence of different stabilities is guided by “invariants”: features that persist across otherwise distinct configurations. While usually they are linked to the (quasi-)material\footnote{The term acknowledges that while certain constraints emerge from the physical or technical properties of a system, others arise from how those properties are interpreted, encoded, or functionally embedded in practice. In postphenomenological accounts, this includes software parameters, interface logics, or data-induced behaviors that shape interaction, even if they are not strictly material in the classical sense \citep{gerlekMaterialityMachinicEmbodiment2025}.}  qualities of technology, more recent interpretations emphasize the role of the environment, in the form of a "rich landscape of affordances" within which the technology is taken up \citep{deboerExplainingMultistabilityPostphenomenology2023}. 

To structure a VCE analysis, \citeauthor{rosenbergerVariationalCrossexaminationMethod2020} suggests identifying features from three broad categories: (A) Networks and Co-shapings—how the technology integrates in sociotechnical networks; (B) Comportments and Habits—how users must physically or cognitively engage with the technology; and (C) Material Tailorings—how the devices are altered to better cater to their contextual goals \citep{rosenbergerVariationalCrossexaminationMethod2020}.

Our analysis builds on a growing body of research that engages with multistability and postphenomenological approaches in the context of digital, data-driven technologies. For example, \citeauthor{lodovicoLeveragingMultistabilityAmbiguity2023}'s work on wearable biosensing shirts demonstrates how ambiguous sensor feedback can encourage multiple stable interpretations of bodily data, depending on the intentions and context \citep{lodovicoLeveragingMultistabilityAmbiguity2023}. In their examination of two projects: a shape-changing bench and digital sticky notes, \citeauthor{jensenPostphenomenologicalMethodHCI2018} use VCE to highlight the utility of the method in Human-Computer Interaction (HCI) context \citep{jensenPostphenomenologicalMethodHCI2018}. In another HCI work, \citeauthor{hauserAnnotatedPortfolioDoing2018} study the human-technology relations through six Research-through-Design (RtD) artifacts, arguing that VCE provides an improved depth of analysis that goes beyond the paradigms of use and utility \citep{hauserAnnotatedPortfolioDoing2018}. Closer to our study, in the domain of generative systems, a recent work has uncovered and compared multiple stabilities of text-to-image models, examining their potential role in the design process \citep{hernandez-ramirezAIEnginesTools2023}.

Specifically in the music domain, explicitly postphenomenological analyses are scarce; however, some works investigate the multistable nature of technical objects in implicit terms. One such example is Beatfield, a multimodal audio-visual system designed with interpretive flexibility in mind, whose multistable nature was evaluated in a qualitative study, contributing a design model that encourages "heterogeneous interpretations of interactive artifacts" \citep{morrealeInfluenceCoauthorshipInterpretation2019}. Moreover, \citeauthor{kieferNalimaMultistableMembrane2024}'s work on Nalima—a membrane-based electro-acoustic feedback instrument—demonstrates how its rich affordances, such as a chaotic feedback system, foster interpretive ambiguity that enables the emergence of diverse stabilities. These stabilities differ in the modes of interaction, the implementation details of the internal feedback mechanism, and the interpretation of the instrument's role \citep{kieferNalimaMultistableMembrane2024}. Finally, Notochord—a low-latency MIDI generation model intentionally designed for creative appropriations—demonstrated its versatility in various roles and contexts \citep{Notochord}. For example, it was used to drive the Magnetic Resonator Piano in an artistic installation \citep{mrp, Armitage2024Augmenting}, while on another occasion, it served as a co-performer playing the Txalaparta, a collaborative music instrument \citep{Txalaparta}.

A close examination of the literature quickly reveals that the concept of multistability can be applied in a rich spectrum of distinct contexts, each of which might operate on a different focal point. As some would argue, this flexible nature of the theoretical concept suggests that multistability should be subjected to its own terms, hence itself becoming a multistable tool \citep{wellnerMultiplicityMultistabilitiesTurning2020}. This remark is reflected in the pivot points, a notion devised by \citeauthor{whyteWhatMultistabilityTheory}, who suggests that the specific type of multistability is determined by the researcher's vantage point \citep{whyteWhatMultistabilityTheory}. For instance, when \citeauthor{kieferNalimaMultistableMembrane2024} describes the distinct stabilities of his instrument Nalima, he takes the instrument itself as a pivot point, and is interested in how its (quasi-)material properties can be exploited in various ways.

In our analysis, \acrshort{gt} itself serves as the pivot point, while our interest lies in how it becomes reconfigured when situated within different artistic contexts.

\section{Analysis using \acrfull{vce}}\label{sec:analysis}

According to \citet{rosenbergerVariationalCrossexaminationMethod2020}, \acrshort{vce} analysis begins by identifying the multiple stabilities of the technology under examination. These stabilities are then cross-examined through the three analytical aspects described earlier: (A) Networks and Co-shapings (Section \ref{coshape}), (B) Comportments and Habits (Section \ref{habits}), and (C) Material Tailorings (Section \ref{tailor}).

This study identifies three principal stabilities of \acrshort{gt} for the purpose of cross-examination:

\begin{itemize}
    \item Stability 1: \acrshort{gt} as an autonomous drum accompaniment generator 
    \item Stability 2: \acrshort{gt} as the rhythmic driver of a harmonic accompaniment generator 
    \item Stability 3: \acrshort{gt} as a generative multi-channel control voltage sequencer in the Eurorack format
\end{itemize}

Stabilities 1 and 3 emerged from the planned development of two concurrent projects. However, Stability 2 was not initially anticipated. Rather, it emerged organically through the experimental use of the system in a new unforeseen context.

\begin{figure}[ht]
  \centering
  \includegraphics[width=1\linewidth]{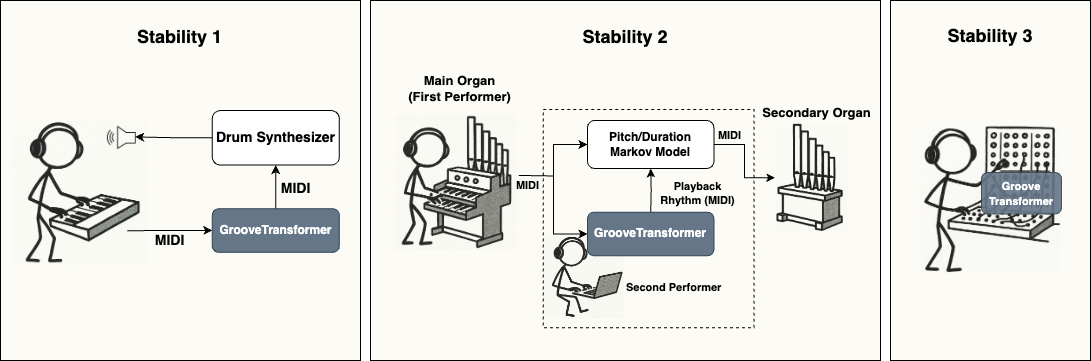}
  \caption{Three perfomance configurations of the system. For Stability 1, Refree improvised on a keyboard synthesizer while \acrshort{gt} rhythmically accompanied autonomously. For Stability 2, Refree improvised on an organ console while a larger system, employing \acrshort{gt} and operated by a second performer, provided harmonic accompaniments on a pair of portable organs. For Stability 3, a Eurorack hardware implementation of \acrshort{gt} was developed and showcased at Sónar+D 2023.}\label{fig:showcases}
\end{figure}


The unintended emergence of the second stability suggests that additional stabilities may be identified through further exploration. However, to ground the present evaluation in observations gathered during the development of the three projects, the following section will limit the examination to the three stabilities identified above.

\subsection{Networks and Co-shapings}\label{coshape}

Here, we examine the system’s role within a broader network of actants, and how the interrelatedness of network's elements and their intentionalities influence its creative possibilities. Recognizing that creative possibility in computational systems is co-constituted, we look beyond performer-system relations to include the dataset and situated context that shape the system. Importantly, the core of the system—a variational transformer-based generative model—has remained invariant between all its described stabilities. The transformer model was trained on a single GrooVAE MIDI dataset \citep{gillickLearningGrooveInverse2019}, created for the sake of a research project by the Google Magenta team. The dataset is comprised of percussive performances of 10 distinct drummers, out of which 80\% are performed by professionals. The dataset features 4/4 metric signature nearly exclusively, and the drummers were asked to enact performances within a list of genres rather limited to the Western standards. As the training data contributes largely to the behaviour of the model, the normativity enacted by the system might have been largely influenced by it.

For the \acrshort{cccb} performance, the system was configured in the role of an autonomous drum accompaniment generator. The performance was set up to test and showcase the capability of the system to act autonomously, in a position of a co-performer, improvising with a professional musician. During the act, the system received a symbolic representation of the keyboard performance and, in response, generated 2-bar drum phrases. Since the system operates in symbolic domain, it was paired with an external VST drum synthesizer with a set of drum instruments selected and tuned before the performance together by the second author and Refree. The expectations from the system were expressed by the performer "to play with him, to force him to go to a place he wouldn't go alone". This hints at the dominant relation to be—in Ihde's terms—the alterity relation\footnote{In postphenomenological theory, Don Ihde distinguishes between several types of human-technology relations. In a \textit{background relation}, technologies recede from immediate attention while still shaping experience (e.g., ambient systems or infrastructure). An \textit{alterity relation} treats the technology as an active “other” with which the human interacts, often framed as quasi-autonomous or co-performing. A \textit{hermeneutic relation} involves interpreting the world through the technology, such as reading a measurement or interacting through a GUI. These relational modes are not mutually exclusive and can co-exist or shift depending on context \citep{ihdeExperimentalPhenomenologyMultistabilities2012}.}: the performer treated the system as an autonomous co-performer on equal musical terms \citep{ihdeExperimentalPhenomenologyMultistabilities2012}, expecting musical agency and initiative, and assuming the uncertainty of its decisions. However, due to the system’s lack of anticipation capacity, the performer reported feeling as though he was "conducing the drums", having to guide the flow of interaction, rather than co-evolve it, assuming a leading role in the performance. Furthermore, since the model's dataset was originally developed for drum pattern generation and used for the same purpose in our system, it reinforced the stylistic norms of Western drumming embedded in the training data.

In the \acrshort{guell} performance, the system took on a new stability and fulfilled a somewhat secondary role. The focus of the show was shifted towards its context—a historic chapel featuring three distinct organ instruments: the main organ, played by Refree; one smaller organ, played by the system as an accompaniment; and a second smaller organ, serving as a system output monitor for the musician. The aesthetic quality of the results was a prevalent intention for the outcomes, and the intentions behind the use of the system were to repurpose the initially drum-oriented technology for the harmony-directed situation. The intentionality of the system here was expressed through a larger network of adjacent technologies, specifically a harmonic generator that assigns pitch to the generated rhythms, and MIDI-controlled acoustic organs that produced the sound. Additionally, as a result of the expressive limitations of the system identified during the first performance, a second performer was employed for the control of both the examined system and the aforementioned harmony generator. The creative agency was therefore shared between both the two performers and the two generative components of the system that mediated the performers' live collaboration. In this sense, the system assumed a number of different roles. From the perspective of the main performer, who played keys, the system assumed a background role—the performer focused on playing the main organ console, while the system provided accompaniment on a pair of portable secondary organs. Since the main performer did not interact with the system directly, and his engagement was mediated through a second performer, the system's autonomy was not foregrounded to him in the same way as in case of the \acrshort{cccb} performance. On the other hand, the second performer approached the system partially as a tool, and partially as an agent. Being able to control the generation through interpretation and live modification of generation parameters, he entered a hermeneutic relation with the system. At the same time, a part of creative agency was delegated to the system, therefore producing an alterity relation. Interestingly, through the repurposing of the rhythmic generator trained on drum data, a creative subversion of the normativity produced by Western drumming tradition takes place. Furthermore, while \acrshort{gt}'s rhythms continued to reflect the tendencies of its original dataset, the addition of the Markov model—trained in real-time on the tonal features of the ongoing performance—helped foreground the performer’s agency. By grounding part of the system in data collected in real-time from the performer, the influence of the external dataset is further diluted while the system becomes more closely aligned with the musician’s expressive trajectory.

Lastly, the Eurorack stability constitutes a special case across all the discussed categories of the analysis. Its special status is derived in a twofold manner: on one hand, its development was not situated within a singular artistic situation, like in the other two cases, but rather positioned in a broader sociotechnical network of modular synthesis; on the other hand, a faculty certainly derived from the modular synthesis ethos, its role is very loosely (un)determined. The lack of situated context directs our interpretation towards the flexibility of use, leading to what could be termed meta-multistability: a stability flexible enough to expose highly multistable nature. Although the system has a tendency to be interpreted as a sequencer module by the user, the way it finally presents itself depends on a flexible context within which it is taken up. Its place within a larger modular system, between accompanying modules and alongside a variety of artistic situations, allow it to assume nearly any musical role. Additionally, the rhythmic biases of the dataset can be repurposed as expressive control signals for other parts of the modular system. This scenario serves as an ultimate expression and manifestation of the underlying system's multi-stability.

\subsection{Comportments and Habits} \label{habits}

A user’s engagement with a given technology is mediated by a set of embodied habits—gestural, perceptual, and cognitive tendencies formed through accumulated experience. As \citeauthor{rosenbergerVariationalCrossexaminationMethod2020} explains, these habits shape how a user comports themselves in relation to the device in order to access a particular stability \citep{rosenbergerVariationalCrossexaminationMethod2020}.

In the case of the \acrshort{cccb} performance, \acrshort{gt} demonstrated its stability as an autonomous drum accompaniment generator, while the performer, Refree, improvised on the keyboard. The modes of interaction were two-fold: the performer could passively or intentionally condition the system through the manner in which they played the keyboard, or they could directly interact with the system's interface for a more immediate response.

The performer, being an experienced musician, preferred to interact with the system in the same manner as he would with a human co-performer—through live improvisation on the keyboard. Drawing on his understanding of how note velocity and micro-timing contribute to the feel of a groove—and knowing that the underlying generative model is conditioned by these parameters—he tended to emphasize the rhythmic qualities of his keyboard performance to coax new patterns and rhythmic variations from the system. At the same time, he was compelled to resist certain habits and comport his playing to two system constraints in order to remain synchronized: a static tempo-set manually by the performer-and a fixed 4/4 time signature. Despite the performer's experience with a range of musical styles, the rigidity of the time signature—combined with the stylistic bias encoded in the model’s training data and the use of a keyboard as the primary interface—invited him to comport his playing style to a Western music tradition. 

In the context of the \acrshort{guell} performance, we consider \acrshort{gt}'s stability as the rhythmic driver for a harmonic accompaniment generation system on pipe organ. Although the system now takes on a more auxiliary role within a larger system, some aspects of the performance remained unchanged: a static 4/4 time signature was necessary to remain synchronized with the system, the performer preferred to interact with the system through keyboard improvisation, and the model's training data had the same stylistic biases. However, there was now a second performer that had control over the system's parameters and that tapped the tempo in real-time based on the main performer. Additionally, in this new context, the performer placed less emphasis on comporting his play style to strictly rhythmic terms. Although his performance still conditioned the rhythmic output of \acrshort{gt}, these outputs were used to drive a harmonic accompaniment, shifting the comportment of his play style from percussive interplay toward a focus on tonal qualities and harmonic progression. 

The stability of \acrshort{gt} operating as a control voltage (CV) sequencer in the Eurorack format presents a much different pattern of comportments and habits compared to the previous stabilities. Here, the performer is unlikely to condition the system through keyboard improvisation; instead, they are interacting with the system directly through a dedicated hardware interface. A performer that has prior experience with DJ mixers, drum machines, or sequencers is likely to interact with the system in a similar manner—rhythmic manipulation of the controls (pattern interpolation, voice density) to shape an improvised composition. This habit is further reinforced by the fact that the most noticeable and immediate interface element is a cross-fader. Additionally, in the Eurorack format, the user is forced to define the musical semantics of the generated patterns. The generated CV sequences are only percussive patterns if the performer decides to patch them as such. In a sense, this highly flexible framework of Eurorack constitutes the \emph{habit of anti-habit}: a comportment oriented toward continual experimentation rather than repetition.

\subsection{Material Tailorings}\label{tailor}

“Material Tailoring” here refers to (quasi-)material adaptations made to the system, both during development and as post-design customizations shaped by specific project requirements.

For the \acrshort{cccb} performance, \acrshort{gt} was tailored for the stability of autonomous drum accompaniment generator.  A looper-like approach with overdubbing allowed accompaniment generation beyond the model’s original two-bar scope. Additionally, a user-defined or rule-based “lifetime” for each event in the input buffer controlled how much of the previous performance context influenced the generated accompaniments. Several heuristics were implemented to autonomously navigate the triangular rhythm space. Most control mechanisms were presented as simple on–off buttons, enabling quick, intermittent adjustments while preserving \acrshort{gt}’s largely autonomous operation.

For the \acrshort{guell} performance, the tailoring incorporated supplementary components outside the system to allow for playback of accompaniments on the secondary, portable organs. A custom Markov-based Max patch converted the rhythmic output of \acrshort{gt} into pitched notes of varying durations. Rather than running fully autonomously, the system was supervised by a second performer, allowing more expressive possibilities for the main performer.

In the Eurorack context, the tailoring took on a physical form. A large horizontal slider provided swift navigation through preset patterns, and the output was delivered as control voltages. Rhythmic input arrived via two control voltage signals for onset and velocity, with an internal clock generator ensuring synchronization with external modules. Instead of on–off buttons, continuous rotary potentiometers offered more nuanced adjustments to rhythmic streams. Finally, some drum voices were grouped together to reduce the number of control voltage outputs; by not labeling these groupings on the interface, the user was encouraged to interpret the system outputs in their own way.

\section{Discussion}\label{sec:reflections}

In this section, we build on the analysis above. First, in Sections \ref{sec:invar}–\ref{sec:situatedness}, we discuss, respectively, the role of invariants, interdisciplinary collaboration, and situatedness in the emergence of multistability. Finally, in Section \ref{sec:vce_dmi}, we reflect on the viability of \acrfull{vce} in the context of Digital Musical Instrument (DMI) design.

\subsection{The role of invariants in the emergence of multistability}\label{sec:invar}

Invariants are the fundamental characteristics of a system that remain constant across its different stabilities and use contexts. In the case of \acrshort{gt}, we can consider how two of its key invariants are not limiting constraints, but rather foundational features that enable specific affordances and contribute directly to its multistable nature.

\acrshort{gt} is a pitch-agnostic, symbolic-to-symbolic system—an architecture that affords considerable flexibility in both input conditioning and output generation. Its pitch-agnostic nature allows a broad range of input sources to influence the generative process, including any source whose rhythmic structure can be represented symbolically: MIDI instrument performances, control voltage triggers, or even audio, when paired with an onset detection system. By functioning without the need for pitched source material, the system enables a variety of use contexts and potential stabilities that might be inaccessible to systems constrained by tonal requirements.

On the output side, \acrshort{gt} emphasizes rhythmic structure while remaining agnostic to sound generation. Rather than relying on an internal audio model or built-in sound engine, it outputs symbolic rhythms that can be interpreted by a wide range of external devices: synthesizers, samplers, MIDI-enabled acoustic instruments, or even non-audio visual setups. This decoupling of rhythm from sound allows for greater artistic flexibility in how the patterns are rendered, performed, or integrated into broader multimodal contexts. Notably, this invariant is what made possible two of the primary stabilities examined in this paper: \acrshort{gt} as a Eurorack control voltage sequencer, and \acrshort{gt} as the rhythmic driver of a harmonic accompaniment generator. Had the system been designed to generate audio rather than symbolic sequences, these particular stabilities would have been precluded. Looking ahead, we can even begin to imagine how the symbolic output could be repurposed away from discrete rhythmic events. By applying various curve-fitting techniques to the generated output sequences, we can allow for a completely new stability: \acrshort{gt} as a continuous rhythmic modulation generator.

\subsection{The role of interdisciplinary collaboration in the emergence of multistability}\label{sec:reseacher_profiles}

The development of \acrshort{gt} was shaped by four collaborators with diverse backgrounds: a researcher with an engineering focus, a researcher with a background in Eurorack performance, a researcher with expertise in digital instrument design, and a professional musician. This variety of perspectives shaped the evolution of the system across its different stabilities.

For example, the Eurorack stability was driven by the researcher with Eurorack performance experience. Their familiarity with modular workflows and live performance scenarios shaped the design of the hardware interface and real-time control architecture. The autonomous drum accompaniment stability, featured in the \acrshort{cccb} showcase, originated from the initial attempt of the third author to develop a real-time generative system \citep{haki_behzad_RT_AIMC_2022}. However, this attempt, undertaken without any collaborators, prioritized the technical performance of the underlying model and struggled to integrate meaningfully into live performance contexts. In the next attempt, the system evolved into a musically functional tool through iterative testing and refinement in collaboration with the professional musician, whose feedback and performance-oriented perspective proved essential in shaping its expressive potential. Finally, the least anticipated and most context-specific stability—that of the harmonic accompaniment system—emerged as a result of a unique opportunity: an invitation from the professional musician to play a multi-organ performance at \acrshort{guell}. This new stability arose by intersecting the researchers’ curiosity to explore broader integrations and the musician’s desire to adapt the system for harmonic purposes specific to this performance context. Consequently, \acrshort{gt}'s multistabilities are not merely an emergent property of the technology itself, but are actively shaped through divergent priorities, ways of thinking and practices within the team. 

\subsection{The role of situatedness in the emergence of multistability}\label{sec:situatedness}

While working on this paper, we have realized how deeply entangled the stabilities are with the situations within which they emerge. In this context, the multistability we refer to throughout this publication is not simply a faculty of the system in isolation, in the essential meaning; instead, it is an observation that the context in which it is used reconfigures it to yet another stability. In this sense, neither the system nor the context can account for the stabilities alone. It is both components that co-constitute one another. This suggests that multistability, even in the case of a complex computational system, is not merely a technical characteristic. It not only includes technical aspects, but also extends to the interpretations, relations, and practices through which the system is taken up. 

The application of the VCE framework has highlighted the crucial role that real-world deployments played in revealing the system’s mediating function in music-making. Equally important is the opportunity afforded by specific artistic contexts, which proved indispensable for a deeper investigation into the system’s affordances and the range of possible interpretations they enable. We recognize that the particularities of the situations in which the system was deployed significantly shaped its development. Had those contexts been different or completely absent, the resulting system would likely have taken a different form.

These observations lead us to conclude that the capacities of such a system can only be fully understood within concrete, situated contexts. To study its potential, those contexts must be carefully aligned with research objectives. At the same time, the very process of examining a system across diverse situations becomes a productive mode of inquiry that inherently leads to more multistable designs. When creative appropriation is a prominent goal, it is worth employing this approach in the engineering process.

\subsection{A reflection on the viability of VCE in the context of DMI design}\label{sec:vce_dmi}

In the context of our research, the VCE has revealed its primary strength as a descriptive and analytical method. Rather than offering prescriptive guidance or explicit engineering recommendations, VCE enabled a deep theoretical engagement with the system's stabilities and forced a methodological, reflective and interdisciplinary discussion among the researchers.

The retrospective character of our study is likely one of the reasons behind the descriptive nature of the outcomes. Since the analysis was conducted post factum, the process has naturally leaned towards the sense-making and interpretation rather than speculative generation of research directions or modifications to the design. Possibly, if the framework was applied during the development process rather than after it, it would have taken a more actionable form, guiding decisions, or encouraging more speculative exploration of alternative configurations. That said, the framework undoubtedly helped to uncover meaningful differences between the identified stabilities, showing how they emerged through differing configurations, roles, and situated uses of the same technology, reinforcing system's multistable interpretation.

Importantly, VCE's grounding in postphenomenological theory proves valuable beyond its utility for structured comparison. It underscores the relational dimensions of experiencing DMIs, shifting the analytical lens from the system as a static artifact to the ways in which it mediates, co-shapes and is co-shaped by its users, the programs of actions and the contexts. Such a perspective encourages broader and more multidimensional understanding of DMI design: one that acknowledges that instruments stabilize not only through their technical realization but also through their use and interpretation \citep{jorda2004instruments}, and within the context of broader sociotechnical networks in which they are embedded.

\section{Future work and conclusion}\label{sec:conclusions}

The multistability of \acrshort{gt} did not arise from a top-down design process or rigid set of engineering goals. Rather, it emerged through an interdisciplinary process shaped by the artistic contexts in which the system was deployed. While the technical core of the system remained largely static, the situated nature of its use and the contrasting perspectives of its collaborators led to meaningful reconfigurations of its form and function. With backgrounds spanning engineering, modular synthesis, digital instrument design, and live performance, those involved naturally embodied the interdisciplinary nature that the VCE framework seeks to foreground. This diversity prevented a singular vision of the system from taking precedence, instead supporting multiple use cases and perspectives from the outset.

Refree, who was oriented towards improvisation and co-creation with an autonomous system, was especially influential in shaping the system's controls and modes of interactions. His ideas about how the system should behave in a live performance context helped guide early iterations of the system and cement some of the eventual core aspects.

Additionally, the concurrency of two parallel design efforts—one focused on autonomous generative performance, the other on hardware interaction—introduced positive constraints that encouraged hybrid implementations and interface designs that could accommodate multiple stabilities. Without the concurrency of these two parallel projects, it’s possible that early design decisions of a singularly focused project could have prematurely closed off the potential for certain stabilities to emerge later in the process. Although we have applied VCE in a retrospective manner, this is, perhaps, the precise reason it could be a valuable analytical framework during the development stages. Its inclusion prevents the imposition of a singular vision and provides a structured means to document identified stabilities, recognize latent affordances, and encourage creative appropriation.

In future iterations of work, we plan to continue deploying our system to novel artistic contexts as a means of surfacing additional stabilities. This will be paired with a concerted effort to improve documentation practices that may serve to highlight evolving affordances and ensure that future adaptations are grounded in well-documented insights. The long-term goal is not to converge on a single ideal form of the instrument, but to cultivate a multistable system that adapts to its users and contexts.

\section{Acknowledgments} 
This research was funded by (1) the Secretaría de Estado de Digitalización e Inteligencia Artificial, and the European Union-Next Generation EU, under the program Cátedras ENIA 2022. "IA y Música: Cátedra en Inteligencia Artificial y Música" (Reference: TSI-100929-2023-1), (2) the Maria de Maeztu Strategic Research Program (CEX2021-001195-M) and (3) the IMPA Project PID2023-152250OB-I00 funded by MCIU/AEI/10.13039/501100011033/FEDER, UE.

\bibliographystyle{apalike}   
\bibliography{references}  

\end{document}